\documentclass[aoas,MSNbibl,nameyear,dvips]{arximspdf}
\usepackage{dcolumn}
\usepackage{graphicx}

\doi{10.1214/11-AOAS503}
\volume{6}
\issue{1}
\pubyear{2012}
\firstpage{55}
\lastpage{82}

\makeatletter

\newproclaim{ASS}{Assumption}

\newcolumntype{d}[1]{D{.}{.}{#1}}

\newcommand{\nnnull}{\mbox{\fontsize{8.36pt}{10pt}\selectfont{00\ldots0}}}
\newcommand{\nnnulll}{\mbox{\fontsize{7.6pt}{10pt}\selectfont{00\ldots0}}}

\makeatother

\begin{document}
\begin{frontmatter}

\title{Optimal pricing using online auction experiments: A~{P}\'olya tree approach}
\runtitle{Optimal pricing using online auction experiments}

\begin{aug}
\author[A]{\fnms{Edward I.} \snm{George}\ead[label=e2]{edgeorge@wharton.upenn.edu}}
\and
\author[B]{\fnms{Sam K.} \snm{Hui}\corref{}\ead[label=e1]{khui@stern.nyu.edu}}
\runauthor{E. I. George and S. K. Hui}
\affiliation{University of Pennsylvania and New York University}
\address[A]{Department of Statistics\\
University of Pennsylvania\\
446 Jon Huntsman Hall\\
3730 Walnut Str.\\
Philadelphia, Pennsylvania 19104\\
USA\\
\printead{e2}} 
\address[B]{Leonard N. Stern School\\
\quad of Business\\
New York University\\
40 West 4th Street, Tisch 910\\
New York, New York 10012\\
USA\\
\printead{e1}}
\end{aug}

\received{\smonth{12} \syear{2009}}
\revised{\smonth{7} \syear{2011}}

%
\begin{abstract}
We show how a retailer can estimate the optimal price of a new product
using observed transaction prices from online second-price auction
experiments. For this purpose we propose a Bayesian {P}\'olya tree approach
which, given the limited nature of the data, requires a specially
tailored implementation. Avoiding the need for a priori parametric
assumptions, the {P}\'olya tree approach allows for flexible inference of
the valuation distribution, leading to more robust estimation of
optimal price than competing parametric approaches. In collaboration
with an online jewelry retailer, we illustrate how our methodology can
be combined with managerial prior knowledge to estimate the profit
maximizing price of a new jewelry product.
\end{abstract}

%

\pdfkeywords{Bayesian nonparametrics,
Polya tree distribution,
second-price auctions,
internet auctions,
optimal pricing}

\end{frontmatter}

\section{Introduction}\label{sec1}

As internet auctions become increasingly popular, the modeling of
auction data is capturing the attention of marketing researchers
[\citet{Chaetal02}]. For instance, \citet{ParBra05}
developed an integrated model to capture the ``whether, who, when, and
how much'' of bidding behavior; \citet{YaoMel08} proposed a
structural model to describe the buyer and seller behavior in internet
auctions and compute model-based estimates of fee elasticity.
\citet{BraPar07} used a~generalized record-breaking model to predict
observed bids and bid times in internet auctions.

In this article we turn to the use of internet auctions to estimate the
profit-maximizing price of a new product. Toward that end, we utilize
second-price auction experiments to learn about the consumer valuation
distribution of a~population of potential consumers of the focal
product, a distribution that we denote throughout by $F$. By valuation
here we mean the \textit{maximum} price that a consumer would be willing
to pay for the product.\footnote{This valuation is also called the
consumer's reservation price in the economics literature.} Thus, $F$
captures the demand curve, and can readily be used to estimate the
optimal profit-maximizing price. While a variety of methods, for
example, direct elicitation/contingent valuation [\citet{MitCar}],
indirect survey methods [\citet{Bre}] and conjoint analysis
[\citet{GreSri78}], can also be used for demand estimation,
analysis of second-price internet auctions can provide a useful
complementary approach to validate demand estimates with online field
data.

In the literature on demand estimation using auctions, researchers
typically impose specific parametric specifications on consumer
valuation distributions [e.g., \citet{autokey13}, \citet
{ParBra05}, \citet{YaoMel08}]. However, in the setting where a retailer
tries to set an optimal price for a \textit{new} product, it seems
unlikely that retailers would have precise knowledge about the
appropriate parametric form for $F$. Furthermore, the limited nature of
available data from second-price auction experiments makes it
particularly difficult to verify the validity of standard parametric
assumptions (e.g., Gaussian, gamma). As will be seen in Section \ref
{sec4}, if
the standard parametric assumptions are invalid, estimation of the
optimal price will be biased, leading to lower profits for the retailer.

To cope with this problem, we propose a specially tailored Bayesian
nonparametric approach [\citet{autokey14}] based on the highly flexible
{P}\'olya tree distribution [\citet{Fer74}, Lavine (\citeyear
{Lav92}, \citeyear{Lav94})] to infer
$F$ from second-price auctions. By avoiding the need to impose a more
limited parametric form, this flexibility is well suited for learning
about consumer valuation for a new product, in particular, for
estimating the profit maximizing price.

Our approach can be outlined as follows. For a new product, a series of
nonoverlapping, second-price internet auction experiments are conducted.
For each such auction we obtain, using third-party software, the total
number of bidders (who may or may not place a bid)\footnote{As discussed
in Section \ref{sec2}, we treat someone who visits the auctioned
product but does
not place a bid as an unobserved bidder whose maximum valuation is below
the winning bid.} and the final transaction price. As discussed in
Section \ref{sec2}, we treat internet auctions using an IPV
(Independent Private
Value) auction framework [\citet{Vic61}], an assumption that is widely
used in the literature [e.g., \citet{HouReg07}, \citet
{HouWoo06}, \citet{Ras06}, \citet{Unj}]. Under the IPV
framework, together
with reasonable assumptions (discussed later), the final transaction
price of each auction can be considered as equal to the second-highest
valuation among the bidders, plus a small increment.\footnote{The
transaction price is the second-highest bid plus a very small increment
(\$0.01). In this paper, we subtract the small increment from the
transaction price to obtain the second-highest bid (and hence the
second-highest valuation); see, for example, \citet{Unj}.} Thus, each
auction provides us with the second highest order statistic of an i.i.d.
sample of known size (the total number of bidders) from the consumer
valuation distribution $F$.\footnote{Throughout this paper we restrict
attention to multiple auctions where it can be assumed that there is no
dependence across auctions. We believe this assumption is reasonable (as
discussed in more detail in Section \ref{sec5}) when the auctions are
nonoverlapping [which minimizes information spillover across auctions,
e.g., \citet{Bapetal09}, \citet{Haretal},
\citet{JanZha11}], and when the coming auctions are not
pre-announced before the
end of the current auction [which minimizes the opportunity for bidders
to engage in forward-looking behavior, e.g., \citet{Zei06}]. In our
particular application, we consider auctions of jewelry products that
are heavily differentiated (products from one retailer are unlikely to
be available at competitors), further reducing potential dependence
across auctions. As an empirical check, we examined the autocorrelations
of the time series of final prices (with and without adjusting for
number of bidders) and found no autocorrelation coefficients to be
significant.} We then use our proposed approach to formulate and update
a {P}\'olya tree distribution based on these observed second highest order
statistics, thereby obtaining the posterior distribution of $F$.

Updating a {P}\'olya tree distribution using only a set of second highest
order statistics presents an interesting implementation challenge. To
tackle this problem, which to the best of our knowledge has not been
addressed in the literature, we have devised a structured partition
scheme that allows for posterior computation using an inexpensive data
augmented Gibbs sampling algorithm that is similar in spirit to the
approach in \citet{Pad02}.

The remainder of this paper is organized as follows. In Section \ref
{sec2} we
discuss the mechanism of online second-price auctions, present our
assumptions, and argue that the observed transaction price can be
considered as the second highest order statistic of a sample of known
size from the valuation distribution. In Section \ref{sec3} we review the
essentials of the Bayesian {P}\'olya tree approach, and propose a specially
tailored formulation and updating scheme that can be used to draw
inference about a consumer valuation distribution~$F$ using second-price
auction data. In Section \ref{sec4} we present numerical simulations to
illustrate the performance of the proposed method. We then set forth an
empirical application of our model in Section \ref{sec5} to estimate the
valuation distribution and then derive the optimal pricing of a new
jewelry product using actual auction data together with elicited expert
managerial prior information. Finally, Section \ref{sec6} concludes with
discussion and directions for future research.

\section{Second-price auction data}\label{sec2}

In this section we discuss the features of the ascending, second-price
online auction considered in this paper, and argue that the winning bids
of such auctions can be used to estimate the valuation distribution of
potential consumers of the auctioned product. Through an example,
Section \ref{sec21} reviews the mechanism of the second-price online
auction. In\vadjust{\goodbreak}
Section \ref{sec22} we argue that, under suitable assumptions, the
winning bid
of each auction can be considered as the second highest order statistic
of a sample (of size equal to the total number of observed and
unobserved bidders) drawn from the valuation distribution.

\subsection{Ascending second-price auctions}\label{sec21}

Ascending, second-price auctions are the most common form of internet
auctions. In such auctions, the person with the highest bid wins the
item but pays the price of the \textit{second-highest bid}, plus a small
increment (e.g., \$0.01). In the auction application we consider, an
automatic ``proxy bidding'' system is used. Under this system, each user
can, at any time, put in his/her maximum bid, and the system will
automatically increase his/her bid if another bidder puts in a larger
bid that is still below the stated maximum bid. For concreteness, let us
illustrate this proxy bidding system with a hypothetical example.

Suppose bidders A, B, C are bidding on a certain item. Bidder A is
willing to pay \$3 for the item; bidders B and C are willing to pay \$5
and \$10 for the item, respectively. The starting price of the item is
\$0.01.

Suppose A enters the auction first, and bids \$3. The ``current bid''
will stay at \$0.01, and A is the current leader. Next, B bids \$5. Now,
the ``current bid'' is increased to \$3.01 (i.e., A's highest bid, plus
a small increment), and~B becomes the current leader. Finally, C bids
\$10. The current bid is now increased to \$5.01, and C is the current
leader. Assuming that no more bids are received, C is the winner of the
auction, and pays the final transaction price of \$5.01, which is equal
to the amount of the second-highest bid (B's), plus a small increment.
Note that the highest bid of \$10 (C's bid) is always unobserved.

In the above example, all bidders are observed: they all placed a bid
during the auction. This is not true in general. In most cases, some of
the bidders are unobserved, that is, the number of observed bids is
generally smaller than the number of bidders. This is because if a
bidder's willingness to pay is smaller than the ``current bid'' (at the
time when the bidder intends to place a bid), he will not be able to
place a bid. Thus, whether a bidder is observed or not depends on the
timing on which the bidders place their bids. For instance, take the
same set of bidders in the last example (A: \$3; B: \$5; C:~\$10), but
assume that they place their bids in the order
B${}\rightarrow{}$C${}\rightarrow{}$A. In this case, when A enters, he
is unable
to place a bid because the current price (\$5.01) is already higher than
his valuation of the product~(\$3). Thus, A does not bid, and is thus
unobserved. Due to the presence of unobserved bidder(s), the number of
bids (in this example, two) is smaller than the total number of bidders
(in this example, three).

Thus, the sequence of bids alone does not tell us the exact number of
bidders in the auction, as some bidders may be unobserved. This issue of
unobserved bidders creates identification problems [e.g., \citet{Unj}].
To avoid this problem, it is necessary to use an external source of
information to record the\vadjust{\goodbreak} total number of unique bidders who accessed
the auction, whether or not he/she placed a bid. In our empirical
application in Section~\ref{sec5}, the jewelry retailer accomplished
this by
using third-party tracking software.\footnote{The tracking software
records the total number of unique IPs that have accessed our auction.
The assumption here is that the number of unique IPs is equal to the
number of unique bidders. This may not be true if the same person uses
two different computers to view our product page; this limitation can be
resolved in the future if one can track the unique userIDs instead of
the IPs.} Thus, throughout this paper, we assume that the total number
of bidders in each auction is known.

In this paper we focus on internet auctions that can be suitably modeled
with an independent private value (IPV) auction framework [\citet
{Vic61}] as described in Section \ref{sec22} below. The IPV framework
is a common
assumption made in the applied econometrics literature to model internet
auctions [e.g., \citet{HouReg07}, \citet{HouWoo06},
\citet{Ras06}, \citet{Unj}]. In our empirical application in
Section \ref{sec5}, we
learned from the jeweler that most consumers purchase jewelry from
internet auctions for their own consumption, and rarely for resale.
Thus, an IPV framework seems appropriate (albeit empirically
unverifiable\footnote{See, for example, \citet{BoaBorKad10},
\citet{LafVuo96}.}) there---different consumers value jewelry products differently because of their
idiosyncratic preferences.

It is important to note at this point, however, that an IPV assumption
may not be appropriate in other applications. The IPV assumption will be
violated, for instance, if bidders' valuations are influenced by the
other bids seen during the auction, or if bidders are trying to figure
out the market value of the auction product (perhaps with resale in
mind) [\citet{Kle99}]. In such situations, the inference about $F$
made by our proposed methodology (which explicitly assumes IPV auctions)
may be questionable, and the results should be viewed with
caution.\vspace*{-3pt}

\subsection{Transaction price and second highest order statistics}\label{sec22}

According to economic theory [\citet{Vic61}], in a second-price
auction, the dominant strategy for each consumer is to place a bid that
is equal to his/her valuation of the product (i.e., the highest price
he/she is willing to pay for the item). Thus, we make the following
assumption:\vspace*{-2pt}
\renewcommand{\theASS}{\Roman{ASS}}
\begin{ASS}\label{assI}
Each bidder will try to place a bid equal to
his/her valuation of the product at some time before the end of the
auction if the current price has not yet exceeded his/her valuation (in
which case he/she will not place a bid).\vspace*{-2pt}
\end{ASS}

Note that the only assumption made about bidder behavior is that each
bidder will try to bid his/her valuation \textit{before the end of the
auction}; beyond that, \textit{no}\vadjust{\goodbreak} assumptions are made about a bidder's
visitation and bidding behavior \textit{during} the auction.
Specifically, the assumption does not preclude bidders with multiple
visits and/or multiple bids. It allows for the possibility that a bidder
may not want to bid on her first visit, but wait till almost the end of
the auction to place such a bid [i.e., ``sniping'' or last minute
bidding; e.g., \citet{RotOck02}]. Or, that she may want to
place a smaller bid on her first visit, followed by a bid equal to her
valuation by the end of the auction, if the current price is still lower
than her valuation [e.g., multiple bidding behavior, \citet
{OckRot06}]. All of these (and other behaviors) are allowed under Assumption
\ref{assI}.

Under Assumption \ref{assI}, the observed final transaction price can be
considered as equal to the second-highest valuation (plus a small
increment) of all the bidders regardless of the bidder's order of
arrival.\footnote{We assume that there will always be two or more
bidders, which is the case for our empirical application.} This is
because the bidders with the first and second-highest valuations will
always bid, that is, the current price is never higher than their
valuations before they bid, regardless of the order by which other
bidders place their bids [\citet{Unj}].

Similar to the previous literature on auction demand estimation [e.g.,
\citet{Ada07}, \citet{BalMarRic97}, \citet{CanPea},
\citet{Unj}], the following further two assumptions about the
sample of
bidders in each auction allow us to use the observed transaction prices
to make inference about $F$:
\begin{ASS}\label{assII}
The set of bidders (observed or unobserved) in
an auction is an i.i.d. sample from the population of all potential
consumers of the auctioned product.
\end{ASS}
\begin{ASS}\label{assIII}
The set of (mostly unobserved) latent product
valuations for each of these bidders is an i.i.d. sample drawn from the
valuation distribution~$F$.
\end{ASS}

With the addition of these assumptions, the final transaction price
minus the small increment can thus be treated as the second largest
order statistic of an i.i.d. sample from $F$. By conducting a set of
identical, independent auction experiments, we can therefore collect a
set of second highest order statistics and associated sample sizes
(i.e., the total number of bidders, observed or unobserved, in each
auction) from a set of i.i.d. samples from $F$. In Section \ref{sec3}
we describe
how such data can be used to draw inference about~$F$.

Let us conclude this section with a brief discussion of why we only
consider the final transaction price, but not the entire sequence of
``current prices'' for inference about $F$. Unlike the final transaction
price, the sequence of current intermediate prices is dependent on the
order by which bidders submit their bids. Thus, the\vadjust{\goodbreak} second highest
current price, for instance, is not equal to the third highest valuation
in general. To see this, consider the following example with four
bidders with the following valuations: (A: \$3, B: \$5, C: \$10, D:~\$15).
Suppose the bidders place their bids in the order of
A${}\rightarrow{}$C${}\rightarrow{}$D${}\rightarrow{}$B. Here, the
final transaction
price is \$10.01, which is equal to the second-highest valuation (\$10)
plus a small increment. The second highest current price (\$3.01),
however, does not correspond to the third highest valuation (\$5),
because bidder B is unable to bid. Thus, absent strong assumptions on
the process of bid submissions, the sequence of ``current prices''
provides only limited information about $F$. Fortunately, as will be
seen in Section \ref{sec4}, restricting attention to only the second-highest
final bids lead to reasonably accurate inference about the
profit-maximizing price.

\section{Methodology for inference about $F$}\label{sec3}

This section describes our proposed {P}\'olya tree approach to inferring the
valuation distribution $F$ from the second highest order statistics
obtained by second-price auctions as described in Section \ref{sec2}.
We begin by
defining notation in Section \ref{sec31}, and then briefly describe, in Section
\ref{sec32}, a general alternative parametric approach that we use as a
benchmark for later comparisons in Sections \ref{sec4} and \ref{sec5}.
In Section
\ref{sec33} we present our nonparametric {P}\'olya tree approach and its
implementation in detail.

\subsection{The set of second highest order statistics}\label{sec31}

Throughout this article, we use the following notation to denote the
auction data. Let $y_{ij}$ be the valuation of the $j$th bidder ($j =
1,\ldots, N_{i} $) in the $i$th auction ($i = 1,\ldots, M$). Without loss
of generality, we rearrange the consumer indexes so that $y_{iN_i} <
\cdots< y_{i2} < y_{i1}$. Of these valuations, as described above, we
assume that only~$y_{i1}$ and~$y_{i2}$ correspond to actual bids, and
that of these only $y_{i2}$ is observed. Thus, for each auction, we
observe only the second highest valuation $y_{i2}$ and the total number
of bidders $N_{i}$ (observed and unobserved) who viewed the auction. For
convenience, in our later development and again without loss of
generality, we further rearrange the auction indices so that $y_{M2} <
\cdots< y_{22} < y_{12}$. The essential statistical challenge here is to
draw inference about~$F$ based only on this set of second highest order
statistics.

\subsection{A parametric Bayesian approach to infer $F$}\label{sec32}

If an appropriate parametric form for $F$ could be specified, for
example, the family of gamma distributions or the family of
truncated-normal distributions, then implementation of the following
parametric Bayes approach would be straightforward. Letting $\theta$
denote the index of the specified family, the likelihood of $\theta$
given the observed second-price auction data would be directly obtained
as the product of the order statistic $y_{i2}$ densities, namely,
%
%
\begin{equation}\label{equ1}
\prod_{i = 1}^{M} p(y_{i2}|N_{i},\theta) = \prod_{i = 1}^{M}
N_{i}(N_{i} - 1)[ 1 - \Psi(y_{i2}|\theta) ] [\Psi(y_{i2}|\theta) ]^{
N_{i} - 2}\psi(y_{i2}|\theta) ,\hspace*{-22pt}\vadjust{\goodbreak}
\end{equation}
where $\Psi(\cdot)$ and $\psi(\cdot)$ here denote the CDF and PDF of
the parametric form, respectively [\citet{CasBer01}]. The
posterior distribution for $\theta$ could then be obtained by using the
likelihood, implicit in (\ref{equ1}), to update a prior distribution
for~$\theta$. When simple analytical posterior forms were unavailable,
Markov chain Monte Carlo posterior calculation could be used to sample
$\theta$ from the posterior [\citet{RobCas04}].

Despite its clear appeal and straightforward implementation, the
performance of such a parametric approach will rely heavily on the
appropriateness of the assumed parametric family, as will be seen in
Section \ref{sec4}. This could be especially problematic in a new product
setting where prior information would be unavailable for guiding such a
selection, and where data consisting of only second highest order
statistics would offer little guidance for validating any such
selection. To avoid the possible misspecification of a parametric
family, we propose an alternative Bayesian {P}\'olya tree approach below.
As will be seen, this {P}\'olya tree approach completely avoids the use of
(\ref{equ1}).\vspace*{-3pt}

\subsection{\texorpdfstring{A nonparametric Bayesian {P}\'olya tree approach}{A nonparametric
Bayesian Polya tree approach}}\label{sec33}

Our proposed nonparametric Bayesian approach for inference about $F$ is
based on {P}\'olya tree distribution representations [\citet{Fer74},
Lavine (\citeyear{Lav92}, \citeyear{Lav94})], which we
briefly review below in Section \ref{sec331}. In
Section \ref{sec332} we then propose a~suitably tailored {P}\'olya tree prior
formulation for second-price auction data. In Section \ref{sec333} we describe
a fast computational procedure for posterior updating of this
formulation, and in Section \ref{sec334} describe how inferential statistics
based on this output can be obtained.\vspace*{-3pt}

\subsubsection{\texorpdfstring{Overview of the {P}\'olya tree approach}{Overview
of the Polya tree approach}}\label{sec331}

Here we provide a brief review of the {P}\'olya tree model. For more
details, including theoretical results and statistical properties,
readers may refer to \citet{Fer74}, Lavine (\citeyear
{Lav92}, \citeyear{Lav94}), \citet{MauSudWil92},
\citet{MulWal97} and \citet{Waletal99}.

A {P}\'olya tree distribution is a probability distribution on probability
measures, which can be seen as a generalization of the widely used
Dirichlet processes. A~{P}\'olya tree distribution with parameters $\Pi$ and
$A$, denoted $\operatorname{PT}(\Pi, A)$, is determined by a nested binary
recursive partition $\Pi= (B_{0}, B_{1}, B_{00}, B_{01}, \ldots)$ of the
range of $F$, together with a set of hyperparameters $A = (\alpha_{0},
\alpha_{1}, \alpha_{00}, \alpha_{01},\ldots)$ that govern the allocation
of random probabilities to each set of the partition $\Pi$. Indexing the
sets by $\varepsilon= \varepsilon_{1} \cdots\varepsilon_{m}$, where
$\varepsilon_{i} = 0$ or 1, a {P}\'olya tree distribution assigns random
conditional probabilities to the sets such that (i) $p(B_{\varepsilon
0}|B_{\varepsilon} ) = C_{\varepsilon0}$ where each $C_{\varepsilon
0}\sim \operatorname{Be}(\alpha_{\varepsilon0},\alpha_{\varepsilon1})$ is a beta
random variable, (ii) $p(B_{\varepsilon1}|B_{\varepsilon} ) =
C_{\varepsilon1} = 1 - C_{\varepsilon0}$, and (iii) the
$C_{\varepsilon0}$'s are all independent. Thus, under a {P}\'olya tree
distribution $\operatorname{PT}(\Pi, A)$, the probability of any set
$B_{\varepsilon} \in\Pi$ is the random probability
%
%
\begin{equation}\label{equ2}
P(B_{\varepsilon_{1} \cdots\varepsilon_{m}}|A) = \prod_{i = 1}^{m}
C_{\varepsilon_{1} \cdots\varepsilon_{i}}.\vadjust{\goodbreak}
\end{equation}
Now suppose we regard $\operatorname{PT}(\Pi, A)$ as a prior
distribution for
our unknown $F$, that is, suppose we treat $F$ as if it were a
realization of (\ref{equ2}) from $\operatorname{PT}(\Pi, A)$. An appealing
feature of
this formulation is that, given data from $F$, the posterior on $F$ is
then also a {P}\'olya tree distribution, which can be obtained by a
straightforward update of the hyperparameters. More precisely, given an
observation $x$ from $F$, the hyperparameters $A = (\alpha_{0},
\alpha_{1}, \alpha_{00}, \alpha_{01},\ldots)$ of the {P}\'olya tree posterior
on $F$ are updated by
%
%
\begin{equation}\label{equ3}
\alpha_{\varepsilon}|x=\cases{\alpha_{\varepsilon}+1, &\quad if $x\in
B_{\varepsilon}$,\cr
\alpha_{\varepsilon}, &\quad otherwise.}
\end{equation}
Note that (\ref{equ2}) also illustrates how the hyperparameters $A =
(\alpha_{0},
\alpha_{1}, \alpha_{00},\allowbreak \alpha_{01},\ldots)$ control the ``strength'' of
the {P}\'olya tree prior. The larger the $\alpha$'s, the less the influence
of an observation on the underlying beta distribution update.

Going further, it turns out that $\operatorname{PT}(\Pi, A)$ can also
be efficiently updated with only the partial information that $x \in
B_{\varepsilon}$ but not whether $x \in B_{\varepsilon0}$ or $x \in
B_{\varepsilon1}$ [\citet{MulWal97}]. In such cases, it suffices
to update $\alpha_{\varepsilon}$ to $\alpha_{\varepsilon} +1$ but leave
$\alpha_{\varepsilon0}$ and $\alpha_{\varepsilon1}$ unchanged, so that
in effect we only need update the hyperparameters up to the known
resolution of the data. In the next subsection we describe a partition
formulation for $\Pi$ that will allow us to exploit this feature when
updating a {P}\'olya tree prior on $F$ with the partial information
supplied by second-price auction data.

The last essential ingredient for the specification of a {P}\'olya model
$\operatorname{PT}(\Pi,\allowbreak A)$ is the choice of a base measure $H$ over
the range of $F$, which may be considered as a prior estimate of $F$.
For a given partition $\Pi$, $\operatorname{PT}(\Pi, A)$ can then be
centered at $H$ by choosing $A$ via
%
%
\begin{equation}\label{equ4}
\alpha_{\varepsilon} = \gamma_{m} H(B_{\varepsilon} ),
\end{equation}
where $\gamma_{m} > 0$ is a preselected function of the
level\footnote{For example, the set $B_{0100}$ has level $m = 4$.}
(depth) $m \equiv m(\varepsilon)$ of the partition indexed by
$\varepsilon$ [\citet{MulWal97}]. By using $\gamma_{m}$ that
increase with $m$, the influence of the data via (\ref{equ3}) can be
lessened for
the deeper levels of the partition, thereby stabilizing the posterior at
those levels. Indeed, for the choice $\gamma_{m} = km^{2}$, $F
\sim
\operatorname{PT}
(\Pi, A)$ will be absolutely continuous with probability one, whereas
when $\gamma_{m} \equiv\gamma$ is constant for all~$m, \operatorname
{PT}(\Pi, A)$
reduces to a discrete Dirichlet process [\citet{Fer74},
Lavine (\citeyear{Lav92}, \citeyear{Lav94})].

\subsubsection{\texorpdfstring{Formulating a {P}\'olya tree prior for second highest bid auction
data}{Formulating a Polya tree prior for second highest bid auction
data}}\label{sec332}

The formulation of a {P}\'olya tree prior $\operatorname{PT}(\Pi_{0}, A_{0})$
requires the specifications of a recursive partition $\Pi_{0} = (B_{0},
B_{1}, B_{00}, B_{01}, \ldots)$ and a set of hyperparameters $A_{0} =
(\alpha_{0}, \alpha_{1}, \alpha_{00}, \alpha_{01},\ldots)$ associated
with the sets of the partition. Let us now consider suitable
formulations of $\Pi_{0}$ and $A_{0}$ for the second-price auction data
setup.

\begin{figure}

\includegraphics{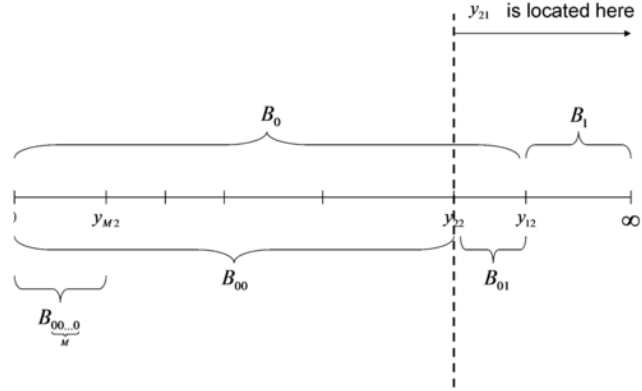}

\caption{The construction of $\Pi_{0}$ for the {P}\'olya tree prior.}\label{fig1}
\end{figure}

We begin with the specification of $\Pi_{0}$, the recursive partition of
the range of $F$ that for our application is $[0,\infty)$. For observed
second highest bid auction data $y_{M2} < \cdots< y_{22} < y_{12}$,
we propose the left-telescoping partition hierarchy $(B_{1} > B_{01} >
\cdots )$ with cut points at the observed $y_{i2}$'s, namely,
%
%
\begin{eqnarray}\label{equ5}
B_{0} &=& (0,y_{12});\qquad B_{1} = [y_{12},\infty);
\nonumber\\
B_{00} &=& (0,y_{22});\qquad B_{01} = [y_{22},y_{12});
\nonumber\\[-8pt]\\[-8pt]
&&\vdots\hphantom{(0,y_{22});\qquad B_{01} =}\vdots
\nonumber\\
B_{00\ldots0} &=& (0,y_{M2});\qquad
B_{00\ldots1} = \bigl[y_{M2},y_{(M - 1)2}\bigr),\nonumber
\end{eqnarray}
depicted graphically in Figure \ref{fig1}. We have formulated this
partition to
facilitate posterior incorporation of all the second-price auction
information in a~computationally efficient manner. This information
consists not only of the observed ordered values of the second-highest
valuations, $y_{M2} < \cdots< y_{22} < y_{12}$, but also includes the
ordering of the unobserved valuations, namely, $y_{ij} < y_{i2}$ $(j > 2)$
and $y_{i2} < y_{i1}$ for each auction $i$. As will be seen in Section
\ref{sec333}, posterior incorporation of the $y_{ij} < y_{i2}$ $(j > 2)$
information with this partition can be done directly through the simple
updating formula (\ref{equ3}), and posterior incorporation of the
$y_{i2} <
y_{i1}$ information can be done with a~multiple imputation scheme based
on a Gibbs sampler. The use of the left-telescoping hierarchy (\ref
{equ5}) is why
imputation is only needed for the $y_{i2} < y_{i1}$ ordering
information. As demonstrated in the Web Appendix I [\citet{GeoHui}], alternative
hierarchies would require the imputation of many more values, vastly
increasing the computational burden of posterior updating.

%
%

Turning to the specification of $A_{0}$ for this partition $\Pi_{0}$, we
propose the use of $\alpha_{\varepsilon} = \gamma_{m} H(B_{\varepsilon}
)$ in (\ref{equ4}) with a base measure $H$ over $[0,\infty)$, which reflects
available prior information. In our empirical example in Section \ref
{sec53}, we
illustrate the elicitation of such an $H$ based on an expert's
subjective judgments. We then consider the corresponding specification
of $A_{0}$ using $\gamma_{m} = km^{2}$ with various values of
$k$. In
the absence of prior information, a~seemingly reasonable default would
be to let $H$ be a uniform distribution over $[0, y^{*}]$, where
$y^{*}$ is the maximum possible valuation of the new
product.\footnote{We recommend and hence assume that $y^{*}$ has been
chosen large enough to be well beyond what anyone would conceivably pay
for the product.} For this $H, H(B_{\varepsilon} )$ would be
proportional to the length of~$B_{\varepsilon}$, when $B_{\varepsilon}$
is bounded. Alternatively, the choice of a proper distribution $H$ with
support $[0,\infty)$ would avoid the need to specify such a $y^{*}$ while
still ensuring that $\alpha_{\varepsilon} = \gamma_{m} H(B_{\varepsilon}
)$ in (\ref{equ4}) would be finite for any $B_{\varepsilon}$.

\subsubsection{\texorpdfstring{Updating the {P}\'olya tree prior given second-price auction
data}{Updating the Polya tree prior given second-price auction
data}}\label{sec333}

Letting $D$ denote our second-price auction data, we are now ready to
describe how our {P}\'olya tree prior $\operatorname{PT}(\Pi_{0}, A_{0})$, with
$\Pi_{0}$ in (\ref{equ5}), can be conveniently updated to obtain the posterior
{P}\'olya tree distribution $\operatorname{PT}(\Pi_{0}, A_{0} |D)$ for~$F$.
Recall that under the assumptions discussed in Section \ref{sec22},
each of the~$M$ second-price auctions is associated with an i.i.d. sample of $N_{i}$
latent valuations $y_{iN_i} < \cdots< y_{i2} < y_{i1}$ from~$F$. Of
these, we only observe the second highest order statistics $y_{M2} <
\cdots< y_{22} < y_{12}$ from each sample. The following update of
$\operatorname{PT}(\Pi_{0}, A_{0})$, based on just this information, is
accomplished by exploiting the particular form of $\Pi_{0}$.

To begin with, the observed second highest bids $y_{M2} < \cdots< y_{22}
< y_{12}$ by definition satisfy $y_{i2} \in[y_{i2},y_{(i - 1)2}) =
B_{\underbrace{\nnnull}_{i - 1}1}$, so that, for $i \ge2$,
%
%
\begin{equation}\label{equ6}
y_{i2} \in B_{\underbrace{\nnnull}_{i - 1}1} \in
B_{\underbrace{\nnnull}_{i - 1}} \in\cdots \in B_{0},
\end{equation}
a consequence of the nesting of the sets in $\Pi_{0}$. Next, although we
do not observe $y_{i3}, \ldots, y_{iN_i}$, we do know that $y_{i2} >
y_{i3}> \cdots> y_{iN_i}$, so that $y_{i3}, \ldots,\allowbreak y_{iN_{i}}
\in(0,y_{i2}) = B_{\underbrace{\nnnull}_{i}}$ and, again because of the
nesting in $\Pi_{0}$,
%
%
\begin{equation}\label{equ7}
y_{iN_{i}},y_{i(N_{i} - 1)}, \ldots,y_{i3} \in
B_{\underbrace{\nnnull}_{i}} \in B_{\underbrace{\nnnull}_{i - 1}} \in\cdots
\in B_{0}.
\end{equation}
Thus, to update the {P}\'olya tree prior for all but the maximum
valuations~$y_{11},\allowbreak y_{21},\ldots,y_{M1}$, we simply increment the $A_{0}$
hyperparameter values\vspace*{2pt} via (\ref{equ3}) as follows. For
each\vspace*{-2pt} auction $i$, we count one value $y_{i2}$ in each of
$B_{0},B_{00},\ldots,\allowbreak B_{\underbrace{\nnnull}_{i -
1}},B_{\underbrace{\nnnull}_{i - 1}1}$ and $(N_{i} - 2)$ values in each
of $B_{0},B_{00},\ldots,B_{\underbrace{\nnnull}_{i}}$.

Beyond the updating above, the only values left to consider are the
maximum valuations $y_{11},y_{21},\ldots,y_{M1}$. Except for $y_{11}$, which
must be located in $[y_{12},\infty) = B_{1}$, there is uncertainty about
the $B_{\varepsilon}$ location of these maximum values. For instance,
consider $y_{21}$; as shown in Figure \ref{fig1}, given that $y_{21} >
y_{22}$ by
definition, we know that $y_{21}$ must be located in either $B_{01}$ or
$B_{1}$, but we do not know which one. What we do know is that each
$y_{i1}$ is located in some~$B_{\varepsilon}$ where the binary index
$\varepsilon$ consists of ($k -1$) $0$'s followed by a~single~1 for some
$k = 1,\ldots, i$. To incorporate this partial information about the
location of $y_{11},y_{21},\ldots,y_{M1}$ into the posterior update of $F$,
we propose a~Gibbs sampler similar to the algorithm proposed by
\citet{Pad02}.

For $i = 1,\ldots, M$, let $z_{i} \in\{1,\ldots, i\}$ where $z_{i} = k
\Rightarrow y_{i1} \in B_{\underbrace{\nnnull}_{k - 1}1}$ indicates the
partition membership of $y_{i1}$. Thus,\vspace*{1pt} the remaining uncertainty about
the update of $A_{0}$ concerns only the unknown values of $Z = (z_{1},
z_{2},\ldots,z_{M})$. Indeed, together with the membership information
in (\ref{equ6}) and (\ref{equ7}), the values of $Z$, if known, would
yield the complete
membership information indicated in Table \ref{tab1}. This information
would then
enable a complete update of $A_{0}$ via (\ref{equ3}), which would in
turn let us
simulate a draw of $C_{\Pi}$, the set of $C_{\varepsilon0}$'s
corresponding to the partition $\Pi$.

%
%
\begin{table}
\caption{The number of observations in each partition, given $z_{i}$'s}
\label{tab1}
\begin{tabular*}{\tablewidth}{@{\extracolsep{\fill}}lccc@{}}
\hline
\textbf{Partition} & \textbf{Count} & \textbf{Partition} & \textbf
{Count}\\
\hline
$B_{0}$ & $\sum_{i = 1}^{M} (N_{i} - 1) + \sum_{i = 1}^{M} I\{ z_{i}
\ge2\} - 1$ & $B_{1}$ & $\sum_{i = 1}^{M} I\{ z_{i} = 1\} + 1$\\[2pt]
$B_{00}$ & $\sum_{i = 2}^{M} (N_{i} - 1) + \sum_{i = 1}^{M} I\{ z_{i}
\ge3\} - 1$ & $B_{01}$ & $\sum_{i = 1}^{M} I\{ z_{i} = 2\} + 1$\\[2pt]
$B_{\underbrace{\nnnulll}_{k}}$ & $\sum_{i = k}^{M} (N_{i} - 1) + \sum_{i
= 1}^{M} I\{ z_{i} \ge k + 1\} - 1$ & $B_{\underbrace{\nnnulll}_{k -
1}1}$ & $\sum_{i = 1}^{M} I\{ z_{i} = k\} + 1$\\[2pt]
$B_{\underbrace{\nnnulll}_{M}}$ & $(N_{M} - 1) - 1$ & $B_{\underbrace
{\nnnulll}_{M - 1}1}$ &
$\sum_{i = 1}^{M} I\{ z_{i} = M\} + 1$\\
\hline
\end{tabular*}
\end{table}

These observations provide the basis for the following Gibbs sampler
updating scheme. First, we simulate $C_{\Pi}$ from $P(C_{\Pi} |A_{0},D,
Z)$, where each~$C_{\varepsilon0}|\allowbreak A_{0}, D, Z\sim \operatorname{Be}(\alpha_{\varepsilon
0}^{D, Z},\alpha_{\varepsilon1}^{D, Z})$ is drawn independently based
on the $(D, Z)$-updated values of~$A_{0}$, namely,
$(\alpha_{\varepsilon0}^{D, Z},\alpha_{\varepsilon1}^{D, Z})$. Second,
conditionally on $C_{\Pi}$, the entries of $Z$ are conditionally
independent.\footnote{This follows immediately from the fact that
conditionally on the realization of $C_{\Pi}$, the probabilities
for the {P}\'olya tree, the $M$ largest bids for each of the samples,
$y_{11},y_{21},\ldots,y_{M1}$, are conditionally independent.} Thus, we
simulate the unknown values of $Z$ from $P(Z|C_{\Pi} )$ which are given
by
%
%
\begin{eqnarray}\label{equ8}
P(z_{i} = 1|C_{\Pi} ) &=& P(y_{i1} \in B_{1}|C_{\Pi} ) = c_{i}C_{1},
\nonumber\\
P(z_{i} = 2|C_{\Pi} ) &=& P(y_{i1} \in B_{01}|C_{\Pi} ) =
c_{i}C_{1}C_{01},
\nonumber\\[-8pt]\\[-8pt]
&&\vdots
\nonumber\\
P(z_{i} = i|C_{\Pi} ) &=& P(y_{i1} \in B_{\underbrace{\nnnull}_{i -
1}1}|C_{\Pi} ) = c_{i}C_{1}C_{01} \cdots C_{\underbrace{\nnnull}_{i -
1}1},\nonumber
\end{eqnarray}
where $c_{i}$ denotes the normalizing constant such that the above
probabilities sum up to 1. This follows directly from (\ref{equ2}) and
the fact
that normalization is needed to account for the membership restrictions
on $y_{i1}$, because our auction data is sorted. By iteratively
simulating from $P(C_{\Pi} |A_{0},D, Z)$ followed by $P(Z|C_{\Pi} )$ in
this manner, this Gibbs sampler can be used to simulate a sequence of
$C_{\Pi}$ that is converging in distribution to $P(C_{\Pi} |A_{0},D)$,
the posterior of $C_{\Pi}$ under $\operatorname{PT}(\Pi_{0}, A_{0} | D)$.

\subsubsection{Inference about $F$}\label{sec334}

It follows from (\ref{equ2}) that under each realization of $C_{\Pi}$
from the
{P}\'olya tree posterior $\operatorname{PT}(\Pi_{0}, A_{0} | D)$, the probability
of a set $B_{\varepsilon_{1} \cdots\varepsilon_{m}}\in\Pi_{0}$ is
given by
%
%
\begin{equation}\label{equ9}
P(B_{\varepsilon_{1} \cdots\varepsilon_{m}}|A_{0},D) = \prod_{i =
1}^{m} C_{\varepsilon_{1} \cdots\varepsilon_{i}}.
\end{equation}
For the purpose of estimating these probabilities, and hence $F$, a
natural estimate in this context is the posterior expectation of (\ref{equ9}),
namely,
%
%
\begin{equation}\label{equ10}
E[ P(B_{\varepsilon_{1} \cdots\varepsilon_{m}}|A_{0},D) ] = E\Biggl[
\prod_{i = 1}^{m} C_{\varepsilon_{1} \cdots\varepsilon_{i}} |A_{0},D\Biggr],
\end{equation}
which we can in turn estimate as follows. Based on a sequence of $T$
draws~from the sequence of $C_{\Pi}$ from the Gibbs sampler (ignoring
$s$ burn-in \mbox{iterations}), we estimate (\ref{equ10}) by the Rao-Blackwellized
version of $\frac{1}{T}\sum_{t = s}^{t = s + T} \prod_{i = 1}^{m}
C_{\varepsilon_{1} \cdots\varepsilon_{i}}^{(t)}$, namely,
%
%
\begin{equation}\label{equ11}
\frac{1}{T}\sum_{t = s}^{t = s + T}\!E\Biggl( \prod_{i = 1}^{m}\!
C_{\varepsilon_{1} \cdots\varepsilon_{i}}^{(t)} |A_{0}, D, Z^{(t)} \Biggr)
\,{=}\,\frac{1}{T}\sum_{t = s}^{t = s + T} \prod_{i = 1}^{m} \frac{\alpha
_{\varepsilon_{1} \cdots\varepsilon_{i}}^{(t)}}{\alpha_{\varepsilon
_{1} \cdots\varepsilon_{i - 1}0}^{(t)}\,{+}\,\alpha_{\varepsilon_{1}
\cdots\varepsilon_{i - 1}1}^{(t)}},\hspace*{-35pt}
\end{equation}
where $\alpha_{\varepsilon_{1} \cdots\varepsilon_{i}}^{(t)}$ is the
updated value of $\alpha_{\varepsilon_{1} \cdots\varepsilon_{i}}$ in
$A_{0}$ based on $D$ and $Z^{(t)}$. This is our posterior estimate of
$F$. The uncertainty of~(\ref{equ11}) as an estimate of~(\ref{equ10}),
due to the unknown
values of $Z = (z_{1}, z_{2},\ldots,z_{M})$, can be summarized by
suitable quantiles of the $T$ values of $\prod_{i = 1}^{m} [
\alpha_{\varepsilon_{1} \cdots\varepsilon
_{i}}^{(t)}/(\alpha_{\varepsilon_{1} \cdots\varepsilon_{i -
1}0}^{(t)} + \alpha_{\varepsilon_{1} \cdots\varepsilon_{i -
1}1}^{(t)}) ]$ appearing\vspace*{2pt} in~(\ref{equ10}). Finally, the uncertainty of
(\ref{equ11}) as an
estimate of (\ref{equ9}) can be summarized by suitable quantiles of the
corresponding $T$ values of~$\prod_{i = 1}^{m} C_{\varepsilon_{1}
\cdots\varepsilon_{i}}^{(t)}$ from the Gibbs sequence.

\section{Simulation study}\label{sec4}

In this section we compare the performance of our proposed {P}\'olya tree
method with Bayesian parametric approaches for estimating
profit-maximizing prices. We consider parametric approaches based on the
gamma and truncated-normal distributions, two parametric distributions
commonly used in marketing research. For the posterior calculation with
these parametric methods, we used a random-walk Metropolis--Hasting
algorithm [\citet{RobCas04}]. We also study the relationship
between sample size and the accuracy of the estimators.

\subsection{Data simulation}\label{sec41}

We conducted three sets of simulation experiments, each using data
simulated from a different functional form for the underlying valuation
distribution $F$. For the data from each $F$, we applied our {P}\'olya tree
approach and the two parametric Bayesian approaches, all using
relatively noninfluential priors, to compute the profit-maximizing
price and the corresponding expected profit. For the {P}\'olya tree prior
$\operatorname{PT}(\Pi_{0}, A_{0})$ with partition $\Pi_{0}$ in (\ref
{equ5}), we set the hyperparameters $A_{0}$ using $\alpha_{\varepsilon}
= km^{2}H(B_{\varepsilon} )$ with $H$ uniform on $[0, y^{*}]$\footnote
{We set $y^{*} = \$20$ here to conform to the bound considered in our
empirical application in Section~\ref{sec53}.} as discussed in Section
\ref {sec332}, with $m = m(\varepsilon)$ denoting the level (depth) of
$B_{\varepsilon}$, and with $k$ set to a small but positive number
$\delta(= e^{-20})$ in order to limit the $\alpha_{\varepsilon}$'s to
being weakly informative. For the $\operatorname{gamma}(a,b)$ and
truncated-normal($\mu,\sigma^{2}$) approaches we used the diffuse
priors $a,b\sim \mbox{truncated-normal}(0,100^{2})$, $\mu\sim
N(0,100^{2})$ and $\sigma\sim\mbox{truncated-normal}(0,100^{2})$.

We evaluate the performance of each method by the expected profit
generated from their estimated profit-maximizing price. First, their
profit maximizing price is obtained by maximizing an estimated expected
(per-bidder) profit function based on the estimate $\hat{F}$ of $F$,
\[
\hat{x} = \mathop{\arg\max}_{x}\hat{\pi} (x) = \mathop{\arg
\max}_{x}\bigl(1 - \hat{F}(x)\bigr)(x - c).
\]
Their corresponding expected (per-bidder) profit is then obtained by
plugging $\hat{x}$ into the actual (``true'') profit function:
\[
\pi(\hat{x}) = \bigl( 1 - F(\hat{x}) \bigr)( \hat{x} - c ).
\]
In each case, the per-unit cost $c$ is taken to be \$5.2 (the actual
per-unit cost for the application in Section \ref{sec5}). Note that the
(per-bidder) profit function is defined by multiplying the proportion of
bidders who have a valuation higher than price $x$ [i.e., $1 - F(x)$] and
the profit for each sale $(x - c)$.

\begin{figure}

\includegraphics{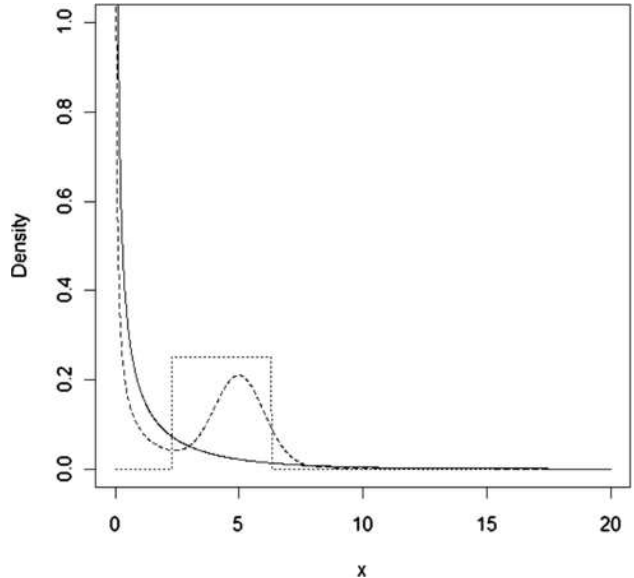}

\caption{``True'' underlying valuation distributions used in the
simulation studies. Solid line: gamma$(0.32, 0.26)$;
broken line: equally
weighted mixture of gamma$(0.32, 0.26)$
and truncated-normal$(5.0, 1.0)$;
dotted line: uniform$(2.3, 6.3)$.}\label{fig2}
\end{figure}

The density functions corresponding to the three underlying $F$
distributions we used are shown in Figure \ref{fig2}. For the first set of
simulations, the underlying $F$ is a gamma distribution with shape
%
%
parameter 0.32 and rate parameter 0.26 (values chosen to replicate
features of the actual data in our empirical application in Section \ref{sec5}).
For the second set of simulations, the underlying $F$ is an equally
weighted mixture of the gamma$(0.32, 0.26)$ and truncated-normal$(5.0, 1.0)$
distributions. From a managerial perspective, this corresponds to a
market with two distinct consumer segments with different average
valuations. From a statistical perspective, this corresponds to a
bimodal distribution for which both of our parameter approaches are
misspecified. For the third set of simulations, the underlying $F$ is
uniform$(2.3, 6.3)$ (centered near the average observed transaction prices
in our empirical application). This is similar to the distribution used
in \citet{JanZha11}.

From each of these three $F$'s, we simulated three data sets containing
$M = 1\mbox{,}000, 100$ and 16 auctions (the number of auctions in our empirical
application). Varying the sample size here sheds light on the
relationship between the sample size and the precision of the
optimal-price and expected profit estimates. For each auction, we
simulated the number of bidders from a Poisson distribution with mean
18.5 (the average number of bidders in our empirical
application).\footnote{We repeated this entire simulation using a
Poisson distribution with mean 37 and found the performance of our {P}\'olya
tree approach to be even better with this larger average number of
bidders.} We then drew the bidders' valuations from $F$, keeping only
the second highest. To account for sample-to-sample variation, we
repeated the $M = 1\mbox{,}000$ case 10 times, and the $M = 100$ and $M
= 16$
cases 100 times, reporting the standard errors along with the mean.

%
%
\begin{table}
\caption{Simulation results. The
standard errors are shown in brackets}
\label{tab2}
\begin{tabular*}{\tablewidth}{@{\extracolsep{4in minus 4in}}lcccccc@{}}
\hline
& \multicolumn{2}{c}{\textbf{{P}\'olya tree}} & \multicolumn{2}{c}{\textbf
{Gamma}} &
\multicolumn{2}{c@{}}{\textbf{Truncated-normal}}\\[-4pt]
& \multicolumn{2}{c}{\hrulefill} & \multicolumn{2}{c}{\hrulefill} &
\multicolumn{2}{c@{}}{\hrulefill}\\
& & \multicolumn{1}{c}{\textbf{Profit}} &
& \multicolumn{1}{c}{\textbf{Profit}} &
& \multicolumn{1}{c@{}}{\textbf{Profit}}\\
& \multicolumn{1}{c}{\textbf{Price}}
& \multicolumn{1}{c}{\textbf{(\$0.01/bidder)}}
& \multicolumn{1}{c}{\textbf{Price}}
& \multicolumn{1}{c}{\textbf{(\$0.01/bidder)}}
& \multicolumn{1}{c}{\textbf{Price}}
& \multicolumn{1}{c@{}}{\textbf{(\$0.01/bidder)}}\\
\hline
\multicolumn{7}{@{}c@{}}{(a) For $\operatorname{gamma}(0.32, 0.27)$
distribution}\\[4pt]
$M=1\mbox{,}000$ & 8.45 & 6.25 & 8.38 & 6.36 & 6.57 & 4.92\\
& (0.22) & (0.04) & (0.02) & (0.00) & (0.01) & (0.01)\\
$M=100$ & 9.13 & 5.72 & 8.29 & 6.33 & 6.54 & 4.86\\
& (0.23) & (0.10) & (0.03) & (0.00) & (0.01) & (0.01)\\
$M=16$ & 8.77 & 5.45 & 8.18 & 6.16 & 6.54 & 4.83\\
& (0.23) & (0.11) & (0.08) & (0.02) & (0.02) & (0.03)\\[8pt]
%
%
%
\multicolumn{7}{@{}c@{}}{(b) For equally weighted mixture between
$\operatorname{gamma}(0.32, 0.27)$ and truncated-normal$(5.0, 1.0)$}\\[4pt]
$M=1\mbox{,}000$ & 5.97 & 8.05 & 6.45 & 6.94 & 6.15 & 7.89\\
& (0.02) & (0.01) & (0.01) & (0.03) & (0.02) & (0.04)\\
$M=100$ & 6.20 & 7.85 & 6.40 & 7.12
& 6.16 & 7.79\\
& (0.13) & (0.09) & (0.01) & (0.03) & (0.02) & (0.04)\\
$M=16$ & 6.49 & 7.16 & 6.41 & 7.04 & 6.29 & 7.19\\
& (0.15) & (0.15) & (0.02) & (0.06) & (0.04) & (0.12)\\[8pt]
%
%
%
\multicolumn{7}{@{}c@{}}{(c) For uniform$(2.3, 6.3)$ distribution}\\[4pt]
$M=1\mbox{,}000$ & 5.77 & 7.53 & 6.31 & 0.13 & 5.68 & 7.43\\
& (0.01) & (0.01) & (0.01) & (0.08) & (0.00) & (0.01)\\
$M=100$ & 5.76 & 7.43 & 6.22 & 2.09 & 5.68 & 7.39\\
& (0.01) & (0.02) & (0.00) & (0.09) & (0.00) & (0.01)\\
$M=16$ & 5.78 & 7.14 & 6.26 & 1.61 & 5.76 & 6.94\\
& (0.01) & (0.06) & (0.01) & (0.17) & (0.03) & (0.16)\\
\hline
\end{tabular*}
\end{table}

\subsection{Simulation results}\label{sec42}

A key feature of our {P}\'olya tree approach is robust estimation of the
profit-maximizing price in the sense that, compared to parametric
methods, it is less sensitive to a misspecified form for the consumer
valuation distribution $F$. Although we would not expect it to perform
as well as a correctly prespecified parametric method, we would like it
to perform better than an incorrectly prespecified parametric method.
Such performance is precisely borne out by our first simulation where
the true $F$ was a gamma distribution. As shown in Table \ref{tab2}(a),
the best
performance was obtained by the gamma parametric approach, for which the
estimated profit-maximizing price was closest to the true value, leading
to the highest expected profit. As expected, the {P}\'olya tree approach
performed slightly worse than the ``correctly specified'' gamma
parametric method but substantially better than the ``incorrectly
specified'' truncated-normal distribution method.

%
%
\begin{table}
\tablewidth=260pt
\caption{Performance of each method (PT, gamma, truncated-normal)
compared to the profit under optimal price}
\label{tab3}
\begin{tabular*}{\tablewidth}{@{\extracolsep{\fill}}ld{3.3}d{3.3}d{3.3}@{}}
\hline
& \multicolumn{1}{l}{\textbf{{P}\'olya tree}} & \multicolumn{1}{c}{\textbf
{Gamma}} &
\multicolumn{1}{c@{}}{\textbf{Trunc-normal}}\\
\hline
\multicolumn{4}{@{}c@{}}{(a) $M=1\mbox{,}000$}\\[4pt]
True: gamma & -1.7\% & 0.0\% & -22.6\%\\
True: mixture & -0.4\% & -14.1\% & -2.4\%\\
True: uniform & -0.4\% & -98.3\% & -1.7\%\\[8pt]
%
%
%
\multicolumn{4}{@{}c@{}}{(b) $M=100$}\\[4pt]
True: gamma & -10.1\% & -0.5\% & -23.6\%\\
True: mixture & -2.8\% & -11.9\% & -3.6\%\\
True: uniform & -1.7\% & -72.4\% & -2.2\%\\[8pt]
%
%
\multicolumn{4}{@{}c@{}}{(c) $M=16$}\\[4pt]
True: gamma & -14.3\% & -3.1\% & -24.1\%\\
True: mixture & -11.4\% & -12.9\% & -11.0\%\\
True: uniform & -5.6\% & -78.7\% & -8.2\%\\
\hline
\end{tabular*}
\end{table}

Turning to the second simulation in Table \ref{tab2}(b), where the true
$F$ was an
equally-weighted mixture of gamma and truncated-normal distributions,
the {P}\'olya tree method performed best in every case except one, where the
size of auctions $M = 16$ was small and the truncated-normal approach
performed slightly better. Finally, for the third simulation in Table
\ref{tab2}(c), when the true $F$ was a uniform distribution, the {P}\'olya
tree method
clearly outperformed both parametric approaches, a situation where the
performance of the gamma approach was particularly bad. Taken together,
the three simulations illustrate how, in contrast to the robustness of
the {P}\'olya tree approach, the parametric approaches can perform poorly
when the parametric form is misspecified.

Table \ref{tab3}(a)--(c) summarizes the results in Table
\ref{tab2}(a)--(c) by comparing the percentage profit loss (compared to
the profit under optimal price), for each method, across the different
values of $M$. As can be seen in \mbox{Table \ref{tab3}(a)--(c)}, the
performance of the {P}\'olya tree method is more robust compared to other
methods, in the sense that it offers the best worst-case performance,
a~minimax kind of appeal. By avoiding the need for a prespecified
functional form, the {P}\'olya tree method avoids the potentially poor
performance due to misspecfication (e.g., using the parametric gamma
method in the third simulation). Finally, with respect to sample size
and estimation accuracy, we note that the estimation accuracy of all
the methods deteriorates with smaller sample sizes $M$. The results in
Tables \ref{tab2} and \ref{tab3} further suggest that if the number of auctions
$M$ is very small (16), it may be helpful to introduce managerial
knowledge through a prior distribution on the valuation distribution.
For that purpose, the {P}\'olya tree approach offers the flexibility of
being able to incorporate prior knowledge by centering the {P}\'olya tree
prior around any base measure $H$, whereas for parametric methods,
prior knowledge is restricted to prior distributions over the
parameters of a particular form.\looseness=-1

\section{Empirical application}\label{sec5}

In this section we apply our method to estimate the profit-maximizing
price of a new jewelry product based on actual data obtained from
second-price auction experiments. In Section \ref{sec51} we describe the
experiments and provide an overview of the data. In Section \ref{sec52}
we apply
and compare our {P}\'olya tree approach with parametric approaches based on
the gamma and truncated-normal distributions. In Section \ref{sec53} we
take a
step further to illustrate the incorporation into our estimation
procedure of a manager's elicited prior beliefs about the consumer
valuation distribution.

\subsection{Data overview}\label{sec51}

In collaboration with an online jewelry retailer, a~total of $M = 16$
identical, nonoverlapping, second-price auction experiments were
conducted on a major internet auction site from February 25, 2006 to
March 20, 2006. Each auction lasted 24 hours, starting and ending at
midnight. The transaction price of the completed auction was recorded
and adjusted for the small increment to obtain the bidders' second
highest valuation $y_{i2}$. Using third-party tracking software, the
jeweler also recorded the total number of unique users who viewed each
auction (i.e., the total number of bidders). The sorted data are shown
in Table \ref{tab4}. To increase the chance of observing some bidding
%
%
\begin{table}
\caption{Sorted data from the sixteen online second-price auction
experiments}
\label{tab4}
\begin{tabular*}{\tablewidth}{@{\extracolsep{\fill}}lcccccccc@{}}
\hline
$N_i$ & 25 & 12 & 22 & 21 & 20 & 27 & 19 & 13 \\
$y_{i2}$ & 10.05 & 8.50 & 5.51 & 5.50 & 5.49 & 5.12 & 4.69 & 4.25 \\
[6pt]
$N_i$ & 19 & 12 & 17 & 22 & 14 & 13 & 25 & 16\\
$y_{i2}$ & 3.73 & 3.53 & 3.25 & 2.34 & 2.26 & 2.02 & 1.50 & 1.25\\
\hline
\end{tabular*}
\end{table}
activity in
each auction, the starting price was always set to \$0.01 with free
shipping. As it turned out, each auction had at least twelve bidders, so
that the second-highest bid was indeed observed in each auction. For the
jewelry product we considered, the per-unit cost $c$ was constant and
equal to \$5.20.

\subsection{Posterior inference for the valuation distribution in the absence
of prior information}\label{sec52}

For the case where prior information was unavailable, we applied the\vadjust{\goodbreak}
methods considered in Section \ref{sec4}, namely, our proposed {P}\'olya
tree method
and the gamma and truncated-normal parametric Bayesian methods with the
weakly informative prior distributions, to the auction data in Table
\ref{tab4}. For the {P}\'olya tree method, we used the partition $\Pi_{0}$
%
%
\begin{table}
\tablewidth=285pt
\caption{First two columns: the partition scheme
$\Pi_{0}$ used in the empirical application. The third column is used
to set $A_{0}$ to approximate the manager's prior beliefs}
\label{tab5}
\begin{tabular*}{\tablewidth}{@{\extracolsep{\fill}}lcc@{}}
\hline
& & \textbf{Prior probability}\\
\textbf{Partition} & \textbf{Interval}
& $\bolds{p_{\varepsilon} =H(B_{\varepsilon} )}$\\
\hline
$B_{0}$ & \hphantom{1}(0.00, 10.05) & 0.901\\
$B_{1}$ & [10.05, $\infty)$\hphantom{00.} & 0.099\\
$B_{00}$ & (0.00, 8.50) & 0.870\\
$B_{01}$ & \hphantom{0}[8.50, 10.05) & 0.031\\
$B_{000}$ & (0.00, 5.51) & 0.731\\
$B_{001}$ & [5.51, 8.50) & 0.139\\
$B_{0000}$ & (0.00, 5.50) & 0.730\\
$B_{0001}$ & [5.50, 5.51) & 0.001\\
$B_{00000}$ & (0.00, 5.49) & 0.729\\
$B_{00001}$ & [5.49, 5.50) & 0.001\\
$B_{000000}$ & (0.00, 5.12) & 0.707\\
$B_{000001}$ & [5.12, 5.49) & 0.022\\
$B_{0000000}$ & (0.00, 4.69) & 0.685\\
$B_{0000001}$ & [4.69, 5.12) & 0.023\\
$B_{00000000}$ & (0.00, 4.25) & 0.663\\
$B_{00000001}$ & [4.25, 4.69) & 0.022\\
$B_{000000000}$ & (0.00, 3.73) & 0.637\\
$B_{000000001}$ & [3.73, 4.25) & 0.026\\
$B_{0000000000}$ & (0.00, 3.53) & 0.627\\
$B_{0000000001}$ & [3.53, 3.73) & 0.010\\
$B_{00000000000}$ & (0.00, 3.25) & 0.613\\
$B_{00000000001}$ & [3.25, 3.53) & 0.014\\
$B_{000000000000}$ & (0.00, 2.34) & 0.534\\
$B_{000000000001}$ & [2.34, 3.25) & 0.079\\
$B_{0000000000000}$ & (0.00, 2.26) & 0.526\\
$B_{0000000000001}$ & [2.26, 2.34) & 0.008\\
$B_{00000000000000}$ & (0.00, 2.02) & 0.502\\
$B_{00000000000001}$ & [2.02, 2.26) & 0.024\\
$B_{000000000000000}$ & (0.00, 1.50) & 0.450\\
$B_{000000000000001}$ & [1.50, 2.02) & 0.052\\
$B_{0000000000000000}$ & (0.00, 1.25) & 0.425\\
$B_{0000000000000001}$ & [1.25, 1.50) & 0.025\\
\hline
\end{tabular*}
\vspace*{-3pt}
\end{table}
in (\ref{equ5}), given by the first two columns of Table \ref{tab5}.
Notice how the
partition elements only split on the leftmost set at each level.

%
%
\begin{figure}

\includegraphics{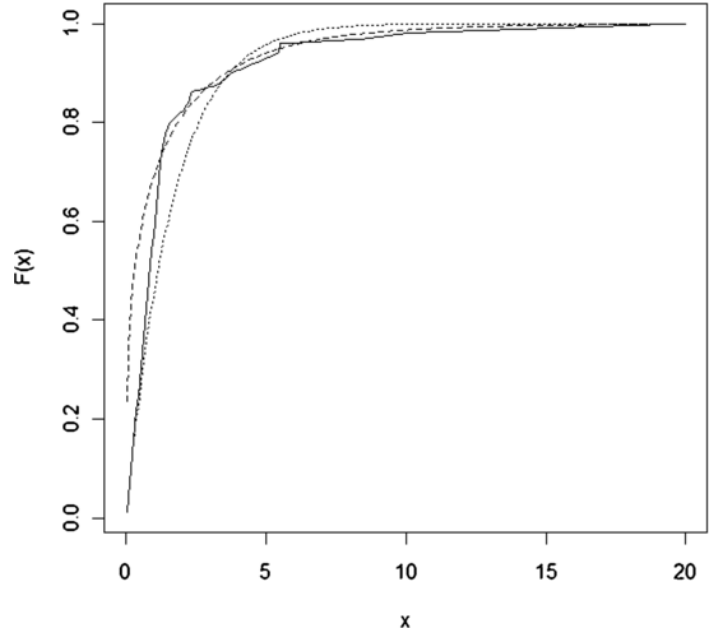}

\caption{Estimates of the valuation distributions $F$ by the three
different methods: {P}\'olya tree (solid line); gamma (broken line);
truncated-normal (dotted line).}\label{fig3}
\end{figure}

\begin{figure}

\includegraphics{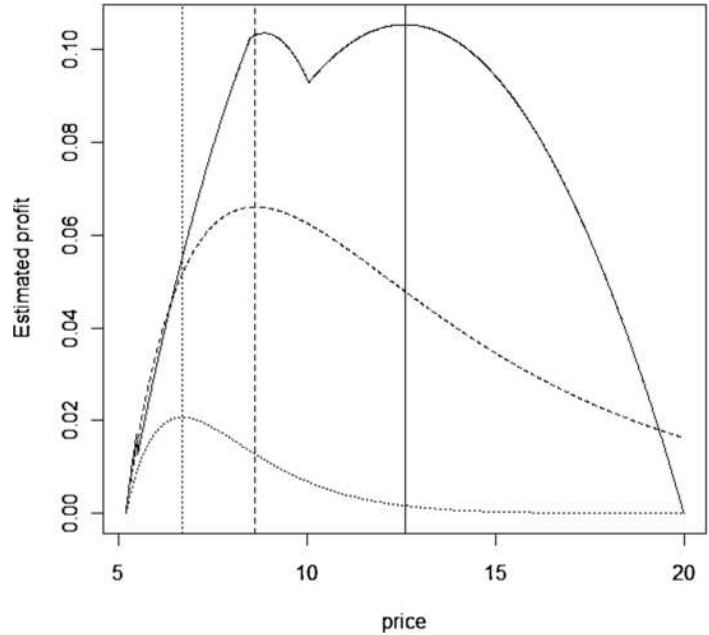}

\caption{Estimated profit functions for the {P}\'olya tree method (solid
line), gamma (broken line); truncated-normal (dotted line).}\label{fig4}
\end{figure}

The estimates of the valuation distribution $F$ for each method are
shown in Figure \ref{fig3},\vadjust{\goodbreak} and the estimated profit functions (along
with the
estimated optimal prices for each method) are shown in Figure \ref
{fig4}. We see
that while the overall shapes of the valuation distributions are quite
similar across all three methods, the quantiles of the three
distributions differ widely. For instance, the median valuation is
\$0.85 for the {P}\'olya tree method, \$0.29 for the gamma method and \$1.13
for the truncated-normal method. Thus, the resulting inference of the
optimal price is similarly highly sensitive to the particular assumption
made for the functional form. The estimated optimal price using the
{P}\'olya tree method is \$12.6, while the estimated optimal prices from
gamma and truncated-normal parametric methods are \$8.63 and \$6.69,
respectively.

\subsection{Incorporating elicited managerial prior beliefs}\label{sec53}

As discussed early, an appealing additional feature of the Bayesian
{P}\'olya tree method is how prior beliefs about $F$ can be
straightforwardly incorporated into the {P}\'olya tree prior $\operatorname{PT}
(\Pi_{0}, A_{0})$. We illustrate this here with the construction of a
prior that incorporates an expert's beliefs about the valuation
distribution $F$ of potential consumers for the auctioned jewelry
product. It is worth noting that it is not clear how to incorporate the
elicited beliefs described below into the parametric priors that we have
been discussing.

In an interview with the manager of the online jewelry retailer behind
our auctions, we used the following subjective CDF construction method
[\citet{Ber85}, page 81] to elicit his prior belief about $F$.
Asking him
to imagine a~hypothetical random sample of 100 consumers, the manager
was asked to state X for various~Y values in the following statement:
``If the price is set at~Y dollars,~X (out of 100) consumers are willing
to buy the product.'' Table~\ref{tab6} shows the set of the manager's responses
[i.e., (X,Y) pairs]. By joining these points with linear segments, these
responses were converted into a cdf, which we denote by
$H$.\footnote{Note that this elicitation method did not capture the
manger's ``uncertainty'' around his prior belief. Future research may
consider how to best capture this uncertainty.}

%
%

%
%
\begin{table}
\tablewidth=260pt
\caption{Manager's prior beliefs about the consumer valuation
distribution}
\label{tab6}
\begin{tabular*}{\tablewidth}{@{\extracolsep{\fill}}d{2.2}d{2.0}@{}}
\hline
& \multicolumn{1}{c@{}}{$\bolds{\ldots}$\textbf{X (out of 100) consumers are}}\\
\multicolumn{1}{@{}l}{\textbf{If the price is set at \$Y}$\bolds{\ldots}$} &
\multicolumn{1}{c@{}}{\textbf{willing to buy the jewelry product}}\\
\hline
0.01 & 98\\
0.50 & 70\\
1.00 & 60\\
2.00 & 50\\
3.00 & 40\\
5.00 & 30\\
7.50 & 15\\
10.00 & 10\\
12.00 & 7\\
14.00 & 6\\
15.00 & 2\\
20.00 & 0\\
\hline
\end{tabular*}
\end{table}

Again using the partition $\Pi_{0}$ in Table \ref{tab5}, we proceeded
to set
$A_{0}$ so that the prior $\operatorname{PT}(\Pi_{0} , A_{0})$ approximates
the manager's prior beliefs. For this purpose, we set
$\alpha_{\varepsilon} = km^{2}p_{\varepsilon}$, the special case of
$\alpha_{\varepsilon} = \gamma_{m} H(B_{\varepsilon} )$ discussed in
Section~\ref{sec332} with $p_{\varepsilon} = H(B_{\varepsilon} )$ and
$m$ the
level of $B_{\varepsilon}$. This setting serves to center the prior at
prior probabilities $p_{\varepsilon} = H(B_{\varepsilon} )$, shown in
the third column of Table~\ref{tab5}, which match the manager's prior
$H$. For
$k$, we considered various values $k = \delta, 10, 20, 50$, to gauge the
effects of different levels of prior uncertainty on the posterior for
$F$.\footnote{As in the simulations in Section \ref{sec4}, we again set
$\delta
= e^{-20}$ to be positive but small.} Larger $k$ reflects a more
certain prior assessment of $F$, yielding a posterior distribution that
is less influenced by the observed data.

\begin{figure}

\includegraphics{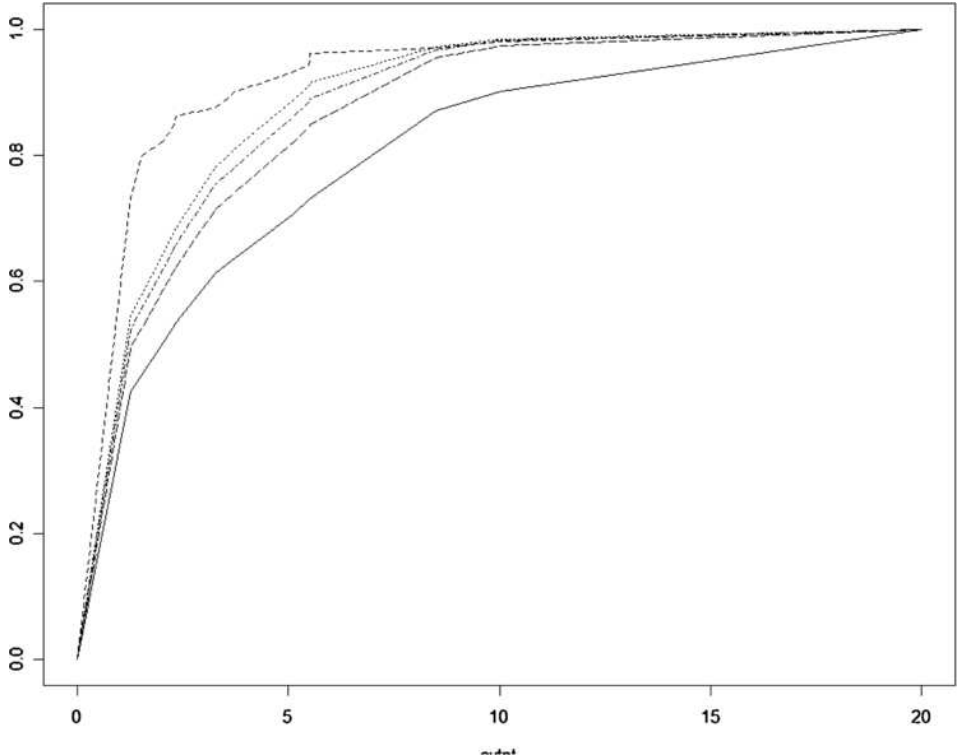}%
\vspace*{-3pt}
\caption{Posterior estimates of the consumer valuation distribution
$F$. The solid line is the manager's prior belief; the other four lines
represent, from top to bottom, the posterior estimates for the four
values $k = \delta, 10, 20, 50$, respectively.}\label{fig5}\vspace*{-3pt}
\end{figure}

For the prior $\operatorname{PT} (\Pi_{0} , A_{0})$ choices described
above, we
estimated the profit-maximizing price. Figure \ref{fig5} shows the various
estimated valuation distributions which incorporate the manager's prior
beliefs. The resulting posterior estimates are shown for the four values
of $k\dvtx \delta$ (top broken line), 10 (second broken line), 20~(third
broken line), and 50 (bottom broken line), along with the manager's
prior beliefs about $F$ (solid line). These results provide a number of
insights. First, as can be seen in the figure, all the posterior
estimates of $F$ are \textit{above} the prior $H$, suggesting that
consumers here have a \textit{stochastically lower} valuation of the
product than that suggested by the manager's prior beliefs. Second, we
observe that with smaller values of $k$, as expected, the posterior
estimate is more influenced by the second-price auction data and less
influenced by the prior.

%
%

\begin{figure}

\includegraphics{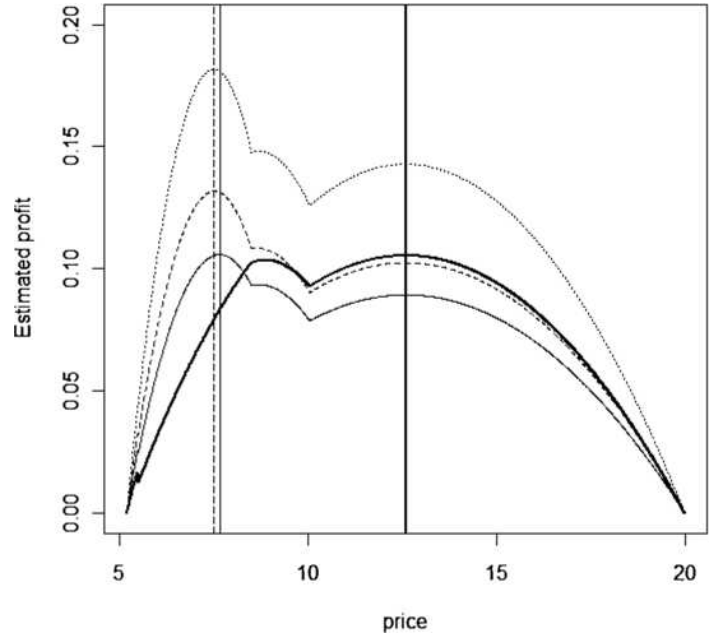}

\caption{Estimated profit functions for the {P}\'olya tree method after
incorporating managerial prior knowledge; $k = \delta$ (thick solid
line), $k = 10$ (thin solid line); $k = 20$ (broken line); $k = 50$
(dotted line).}\label{fig6}\vspace*{-3pt}
\end{figure}

Next, we turn to estimating the profit-maximizing price for each value
of~$k$. The profit function for each value of $k$, along with the
estimated profit maximizing price, is shown in Figure \ref{fig6}.
Figure \ref{fig6} offers some insights about two potential pricing
strategies. There are two price points (around \$7.50 and \$12.60),
that roughly correspond to two pricing strategies commonly used in new
product pricing [e.g., \citet{Tel}]: (i) a ``skimming'' strategy
that targets only a high-value consumer segment (hence achieving very
low volume, but high profit per transaction), and (ii) a
``penetration'' strategy where the retailer sets the price lower in
order to achiever a higher initial penetration, but a lower
profit-per-transaction. The relative effectiveness of each strategy
depends on the value of $k$, that is, the amount of weight that the
manager puts on his prior belief.

%
%

The estimated profit maximizing prices are \$12.6, \$7.66, \$7.52 and
\$7.50 for $k = \delta, 10, 20, 50$, respectively.\vadjust{\goodbreak} We find that for $k <
4$, a skimming strategy is more attractive; for $k > 4$, a penetration
strategy gives better profits. Thus, our method allows the retailer to
quantify and compare the effectiveness of skimming vs. penetration
strategies at any given $k$. Note also that somewhat
counter-intuitively, a stochastically higher valuation distribution
(using larger $k$) here leads to a lower optimal price. Although at each
price a larger percentage of customers will buy the product, the effect
of this on profits is more pronounced at the lower prices.

As can be seen in Figure \ref{fig6}, it appears that by incorporating
some degree
of prior managerial knowledge, the optimal price is estimated to be
around \$7.50. This can be used as a starting point for pricing the new
jewelry product. Based on our recommendations, the jeweler implemented a
fixed price of \$7.49 when the new jewelry product was brought into
market in late 2006.

\begin{figure}

\includegraphics{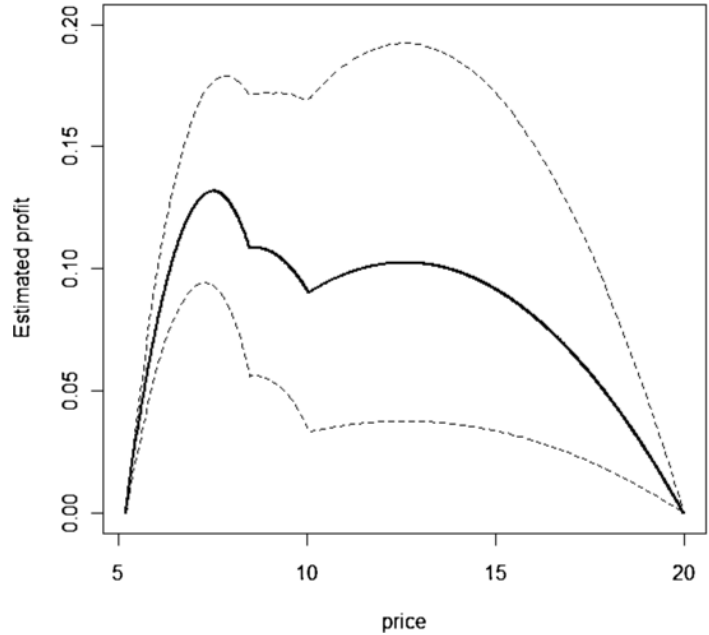}

\caption{Pointwise 90\% posterior intervals of the profit function ($k
= 20$ case).}\label{fig7}
\end{figure}

Our method allows us to not only estimate the profit-maximizing price,
but also to quantify the uncertainty for estimated profits under the
optimal price, by using the posterior sample draws from the {P}\'olya tree.
Figure \ref{fig7} displays\vadjust{\goodbreak} the pointwise 90\% posterior intervals for
the profit
function when $k = 20$, which reflects the degree of uncertainty for our
results. For example, the estimated profit (for the $k = 20$ case) at
the optimal price of \$7.48 is \$0.14 per bidder, with a 90\% posterior
interval of (\$0.09, \$0.18). This provides the retailer with an
estimate of the range of profit that can be obtained.\vspace*{-3pt}

\section{Discussion and future research}\label{sec6}

In this paper we have developed a~nonparametric Bayesian methodology
that enables retailers to estimate the optimal price for a new product
by learning about the consumer valuation distribution from second-price
auction data. Using a flexible {P}\'olya tree distribution to represent
uncertainty about the unknown consumer valuation distribution, we have
proposed a {P}\'olya tree prior formulation and computational approach that
allows for fast updating of the hyperparameters using only
second highest order statistics obtained from a set of auctions. Through
collaboration with an online jewelry retailer, we apply our methodology
to incorporate managerial prior beliefs and derive the optimal price for
a new jewelry product. The generality of our proposed methodology allows
for its application to many different products.

%
%

A key to the computational advantages of our setup is the use of the
observed second order statistics as the cutpoints for the prior
partition $\Pi_{0}$ in~(\ref{equ5}).\vadjust{\goodbreak} Although strict Bayesian coherence is
violated by the use of the data to formulate the prior partition, it
does not seem that the injected structural information is creating a
particular bias.\footnote{Note that we only endorse a data-dependent
partition insofar as the $y_{i2}$'s are used as the cutpoints.
Beyond that, further data-dependent partitions may be ill advised. To
take an extreme example, suppose one introduced the finer partitions
$B_{10} = [y_{12}, y_{12} + \delta]$ and $B_{11} = [y_{12} + \delta,
y^{*}]$. For small enough $\delta$, the resulting posterior would allocate
an inappropriate amount of weight to the very small interval $B_{10}$.}
Nonetheless, because the {P}\'olya tree posterior may still be influenced by
$\Pi_{0}$ in addition to $A_{0}$, it is important to be mindful of the
impact of some of its basic characteristics. While {P}\'olya tree
generalizations involving random partitions [e.g., \citet{Padetal03},
\citet{WonMa10}] would be a way to mitigate this influence,
the computational burdens of their implementation would likely be
overwhelming for the second-price auction data.

One aspect of $\Pi_{0}$ that does appear to incur some systematic bias
is the assignment of the $y_{i2}$'s to the upper intervals (e.g.,
$y_{12} \in B_{1}; y_{22} \in B_{01}$, etc.) by defining the upper
intervals $B_{\varepsilon1}$ in (\ref{equ5}) to be left closed.
However, the
upward bias resulting from posterior updating with this upper interval
assignment is substantially smaller than the downward bias that would
result from a lower interval assignment (details available upon
request). Another alternative, left for future research, might be to
consider partial probabilistic assignments of each of the $y_{i2}$'s to
both intervals.\vadjust{\goodbreak}

Finally, the choice of the left telescoping hierarchy does also
influence the posterior. As illustrated in Web Appendix I [\citet{GeoHui}], this
influence of the chosen hierarchy is lessened when $\alpha_{\varepsilon}
= \gamma_{m} H(B_{\varepsilon} )$ with $\gamma_{m}$ chosen very small,
so that $\gamma_{m}$ is approximately constant, at least at the lower
levels. However, this strategy would be inappropriate for the scenario
in Section \ref{sec53}, where we would not want to minimize the impact
of an
informative managerial prior. Due to the level dependent weighting of
the prior through $m^{2}$, the intervals at the deeper levels have a
stronger prior, resulting in a posterior that will be sensitive to the
choice of hierarchy. In future work, it may be useful to consider
alternative hierarchies that may better represent the manager's prior
beliefs and uncertainty about them. We leave the issue of eliciting the
most reasonable hierarchy and associated level-dependent weighting
function as a future research direction.

To conclude, our research adds to the recent and growing stream of
literature on the use of Bayesian nonparametric techniques in marketing
[e.g., \citet{Braetal06}, \citet{BreSte08}, Kim, Menzefricke
and Feinberg
(\citeyear{JinMenFei04,JinMenFei07}), \citet{SooJamTel09}].
Bayesian nonparametric techniques
provide a rich toolkit that allows modelers to avoid imposing
restrictive parametric functional forms. \citet{Braetal06} and
Kim, Menzefricke and Feinberg
(\citeyear{JinMenFei04}) utilize a Dirichlet process prior to specify the
heterogeneity distribution; \citet{BreSte08} and Kim, Menzefricke
and Feinberg
(\citeyear{JinMenFei07}) use a Bayesian spline approach to model the
price response
function. In the same spirit, this paper introduces the {P}\'olya tree prior
to model uncertainty about an unknown consumer valuation distribution
for the purpose of optimal price estimation. To the best of our
knowledge, this is the first marketing application to make use of a
{P}\'olya tree distribution; we certainly hope that in the future, this
flexible class of distributions will be added to the modeler's toolkit.

\section*{Acknowledgments}

The authors are very grateful
to the reviewers for their generous insights.

\begin{supplement}[id=suppA]
\stitle{Web Appendix for ``Optimal pricing using online auction
experiments: A~{P}\'olya tree approach''}
\slink[doi]{10.1214/11-AOAS503SUPP} 
\slink[url]{http://lib.stat.cmu.edu/aoas/503/supplement.pdf}
\sdatatype{.pdf}
\sdescription{Robustness checks for the left telescoping hierarchy and
the IPV
assumption can be found in the supplemental article.}
\end{supplement}

%
%

\printaddresses

\end{document}